\documentclass[conference]{IEEEtran}
\IEEEoverridecommandlockouts
\usepackage{cite}
\usepackage{amsmath,amssymb,amsfonts}
\usepackage{algorithmic}
\usepackage{graphicx}
\usepackage{caption}
\usepackage{textcomp}
\usepackage{xcolor}
\usepackage{svg}
\usepackage{afterpage}
\usepackage{booktabs}
\usepackage{float}
\usepackage{subfigure}
\usepackage{multirow}
\usepackage{todonotes}
\usepackage{soul}
\usepackage[citecolor=blue,colorlinks=true,linkcolor=blue]{hyperref}%

\def\BibTeX{{\rm B\kern-.05em{\sc i\kern-.025em b}\kern-.08em
    T\kern-.1667em\lower.7ex\hbox{E}\kern-.125emX}}
\begin{document}

\title{A Lightweight FPGA-based IDS-ECU Architecture for Automotive CAN}

\author{\IEEEauthorblockN{Shashwat Khandelwal \& Shanker Shreejith}
\IEEEauthorblockA{Department of Electronic \& Electrical Engineering,
Trinity College Dublin\\
Dublin, Ireland\\
Email: \{khandels, shankers\}@tcd.ie}}



\maketitle

\begin{abstract}
Recent years have seen an exponential rise in complex software-driven functionality in vehicles, leading to a rising number of electronic control units (ECUs), network capabilities, and interfaces. 
These expanded capabilities also bring-in new planes of vulnerabilities making intrusion detection and management a critical capability; however, this can often result in more ECUs and network elements due to the high computational overheads.  
In this paper, we present a consolidated ECU architecture incorporating an Intrusion Detection System (IDS) for Automotive Controller Area Network (CAN) along with traditional ECU functionality on an off-the-shelf hybrid FPGA device, with near-zero overhead for the ECU functionality. 
We propose two quantised multi-layer perceptrons (QMLP's) as isolated IDSs for detecting a range of attack vectors including Denial-of-Service, Fuzzing and Spoofing, which are accelerated using off-the-shelf deep-learning processing unit (DPU) IP block from Xilinx, operating fully transparently to the software on the ECU.
The proposed models achieve the state-of-the-art classification accuracy for all the attacks, while we observed a 15$\times$ reduction in power consumption when compared against the GPU-based implementation of the same models quantised using Nvidia libraries. 
We also achieved a 2.3$\times$ speed up in per-message processing latency (at 0.24\,ms from the arrival of a CAN message) to meet the strict end-to-end latency on critical CAN nodes and a 2.6$\times$ reduction in power consumption for inference when compared to the state-of-the-art IDS models on embedded IDS and loosely coupled IDS accelerators (GPUs) discussed in the literature.
\end{abstract}

\section{Introduction}\label{sec:introduction}

Automotive networks are continually increasing in size and complexity as advanced electronic systems like driver assistance systems (ADAS) and high-bandwidth sensors are becoming increasingly common on most production vehicles.  
Furthermore, such systems are offering increased connectivity to the outside world to enable new waves of safety and comfort applications.
Such increased connectivity of various electronic control units (ECUs) (and the vehicular networks these communicate over) allows remote monitoring and control of critical systems for diagnostics, software services and over-the-air upgrades. 
However, while they enable constant upgrade and upkeep of vehicles to improve their longevity, such interfaces also opened new avenues for attackers to deploy both invasive and non-invasive schemes to inject malicious content on these previously siloed internal networks~\cite{nie2017free,iehira2018spoofing,cai20190}.
Attacks that involve taking control of the vehicular network leading to control its critical functions have been demonstrated by multiple research groups~\cite{greenberg2015after}.
These attacks have been made possible because of the inherent lack of security features in legacy networks like Controller Area Network (CAN)~\cite{CanBosch} that are used in vehicular networks for critical information exchange among ECUs.
With wireless networks and vehicle-to-X communication becoming increasingly common, more pathways emerge for attackers to launch multi-vehicle attacks without requiring physical access or tampered ECUs. 

Multiple mitigation strategies have been proposed to address these challenges, prominent among which are intrusion detection systems (IDSs). 
These systems integrate as an ECU on the network and analyse the flow of network traffic including the message contents in vehicle networks looking for unusual network activities or discrepancies.
Any unusual behaviour is flagged as a possible threat, the severity of which is evaluated by critical systems to determine appropriate actions ranging from alerting the user to falling back to a minimal `safe operating mode'. 
Early IDSs relied on rules/specifications ~\cite{larson2008approach,miller2013adventures} that captured parameters of network and messages under standard operating mode or detected signatures of previously known attacks based~\cite{studnia2018language} to determine unusual activities. 
However, these IDSs suffered from multiple issues including higher false positive and false negative detection, inability to scale to newer attack and memory overheads for each new rule or signature. 
Recently, machine learning (ML) based IDSs have shown as a viable alternative with their higher detection accuracy, adaptability to newer attack modes and their generalisable nature~\cite{seo2018gids,song2020vehicle,agrawal2022novelads,cheng2022tcan,ma2022gru,desta2020mlids}.
Despite the performance, their deployment in electric/electronic (E/E) systems continues to be non-trivial due to a combination of complex multi-standard network architecture (automotive Ethernet, CAN, FlexRay and other network protocols), complex ML models for IDS with provisions for over-the-air updates, tight power budget restricting the use of powerful GPUs for near-line-rate detection of threats and the weight/cabling overheads due to the preferred integration architecture of standalone IDS ECUs. 
The lack of functional consolidation stems from the inability to achieve clean resource partition between (critical) tasks on automotive (multicore) processor-based ECUs; however, loosely coupled accelerators (like GPUs) have been explored to cater to the complexity of adaptive systems like ADAS although these incur much higher power consumption and interfacing complexities. 


Alternatively, ECU architectures based on hybrid FPGAs have shown the ability to achieve clear isolation between consolidated tasks on the same die, while also allowing specialised accelerators to be closely coupled to the functions improving their performance and energy efficiency. 
On a hybrid FPGA, a specialised hardware-efficient accelerator can accelerate the  ML-IDS in isolation, while the capable ARM cores (with any custom accelerator) performs the ECU function on the same die, allowing seamless integration of distributed IDS capabilities.  
Prior research has explored the case for (hybrid) FPGA-based ECUs to enable compute acceleration of complex tasks in vehicular systems~\cite{cho2021fpga,shreejith2013reconfigurable,zcu104link}, functional consolidation~\cite{vipin2014mapping} and reliability~\cite{BoschMS6}.

In this paper, we define an ECU architecture for enabling distributed IDS in vehicular networks with the IDS function deployed in isolation from the main function while retaining software control of its execution. 
Further, we explore a light-weight feed-forward ML model for IDS and deploy them through off-the-shelf deep-learning accelerator (Xilinx DPU) on a Zynq Ultrascale+ hybrid FPGA platform. 
The key contributions are as follows:
\begin{itemize}
    \item We present a custom 8-bit feed-forward quantised multi-layer-perceptron (QMLP) based-IDS for automotive CAN achieving state-of-the-art classification accuracy across multiple attack vectors, all of which are detected using a single IDS ECU.  
    \item The IDS-integrated ECU architecture where the IDS is deployed using off-the-shelf Xilinx DPUs modelling an AUTOSAR compliant architecture while allowing fully isolated execution of IDS task.
    \item The tightly integrated approach achieves noted improvements in terms of per-message processing latency and power consumption against the state-of-the-art IDSs in literature and when the quantised version of the model is implemented using loosely coupled accelerators like a GPU.
\end{itemize}
We evaluate our model and integration efficiency using the open CAR Hacking dataset with the entire CAN data frame used as an input feature to improve the detection performance. 
Our experiments show that the proposed lightweight QMLP-IDS achieves an average accuracy of 99.96\% across multiple attack vectors such as Denial of Service (DoS), Fuzzy, and spoofing (RPM and Gear) attacks on a single device, identical to or exceeding the detection accuracy achieved by state-of-the-art \text{GPU- and CPU-based} implementations. 
The tightly integrated ECU architecture reduces the per message execution latency by 2.3$\times$ and the power consumed by 2.6$\times$ compared to state-of-the-art IDSs proposed in the research literature. 
We also see a reduction of 15$\times$ in power consumption when the quantised version of the IDS is implemented on a GPU.

The remainder of the paper is organised as follows. Section~\ref{sec:background} provides background information on the CAN protocol, IDS approaches and Quantised Neural Networks; section~\ref{sec:proposedmodel} describes the proposed MLP model and the implemented multi-core architecture on the Ultrascale+ device; section~\ref{sec:experiments} outlines the experiment setup and results; and we conclude the paper in section~\ref{sec:conclusion}. 

\section{Background and Related Works}\label{sec:background}
\subsection{Controller Area Network}
\begin{figure}[t!]
  \centering
  \includegraphics[width=\linewidth]{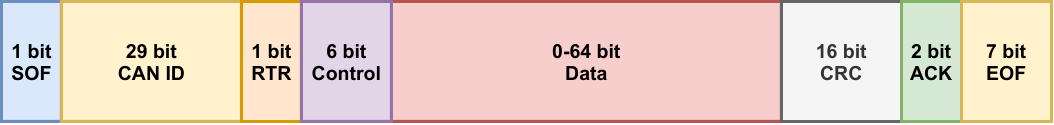}
  \caption{Frame format of an extended frame CAN message that uses a 29-bit CAN ID.}
  \label{fig:1}\vspace{-5mm}
\end{figure}

In-vehicle networks enable distributed ECUs to exchange control and data messages to achieve the global functions of the vehicle. 
Multiple protocols are used in vehicular systems to cater to different functions based on their criticality and to optimise the cost of E/E systems. 
CAN~\cite{CanBosch} and CAN-FD~\cite{hartwich2012can} continues to be the most widely used protocol today due to its lower cost, flexibility and robustness. 
Figure~\ref{fig:1} illustrates the bit-field definition of a CAN data frame, with each segment providing some function in the network operation. 
Each bit-field within the CAN frame serves a specific purpose: start of frame (SOF) bit indicating a start of a new message, remote transmission request (RTR) bit to request information from another ECU, acknowledge (ACK) field to indicate errors in frame and the cyclic redundancy check (CRC)  field carrying a CRC value. 
The CAN ID (11-bit base, 29-bit extended format) is a unique identifier assigned to each message and by extension, defines its priority for transmission on the shared bus.
The message itself is contained in the data field.

The broadcast CAN bus uses a bit-wise arbitration method to control medium access to the bus using the CAN ID allocated to each message. 
CAN also supports multiple data rates (125\,Kbps to 1\,Mpbs) and multiple modes of operation (1-wire, 2-wire) to cater to a range of critical and non-critical functions in vehicles. 
Despite this robustness, CAN is inherently insecure: there is no built-in mechanism in the network to authenticate the transmitter, receiver, or the message content itself~\cite{8658720}.  
This makes CAN vulnerable to simple and efficient attacks like message sniffing, fuzzing, spoofing, replay attacks, and Denial of Service (DoS) attacks~\cite{mukherjee2016practical,enev2016automobile,koscher2010experimental,palanca2017stealth}. 
These attacks rely on physical access to the CAN network; hence onset of such attacks indicate compromised ECUs on the critical network and thus detecting such intrusions are of key interest. 

\subsection{IDSs for CAN}
Researchers have explored various techniques to detect intrusions in CAN bus.
These can be broadly classified into flow-based, payload-based, and hybrid schemes that combine the flow-based and payload-based techniques~\cite{al2019intrusion}.
Flow-based approaches extract identifiable traits like message frequency and/or interval for the network and use these to determine abnormal activity on the network~\cite{vuong2015performance}.
Payload-based schemes use the information in the CAN frames to detect abnormal sequences of instructions or data~\cite{narayanan2015using}.
Hybrid scheme uses both the timing information and the CAN frame fields to create a more holistic view, allowing them to extract specific signatures of transmitting ECUs, receiving ECUs, and messages~\cite{weber2018embedded,vasistha2017detecting}. 
For instance, the fingerprint-based approach uses low-level electrical signal levels and timing of signals in relation with the message contents to identify potential intrusions when large deviations are observed~\cite{cho2016fingerprinting,cho2017viden}.
Parameter monitoring-based techniques extract message-level frequency and timing characteristics for different fields or specific messages, like offset ratio and timing in case of a remote transmission frame, to identify potential intrusions~\cite{lee2017otids}.
Researchers have also proposed the use of entropy of the network traffic as a method to identify abnormal conditions and intrusions~\cite{wu2018sliding}.
Such approaches are further generalised by machine learning based-techniques, including classification approaches, deep learning-based schemes and sequential techniques.

\subsection{ML-based IDS}
Most ML-based approaches fall under the hybrid scheme, utilizing both the content of the message frame and their timing/frequency characteristics to achieve better detection. 
In~\cite{alshammari2018classification}, the authors use the offset ratio and time intervals between remote requests to classify DoS and fuzzy attacks using a Support Vector Machine classifier. 
Tree-based approaches explored by~\cite{yang2019tree} use CAN IDs and data segment contents as input features to tree-based classifiers like decision trees, random forests and others. 
In~\cite{song2020vehicle}, the authors propose a reduced inception net architecture for IDS which uses deep convolutional neural networks. 
The authors show that the ML architecture can achieve over 99\% accuracy across DoS, fuzzing, and spoofing attacks.
The authors used a dataset captured from the actual vehicle for training and testing their model, which has been shared with the community for further research.
Since the dataset covers multiple attack modes with actual CAN messages, we use the same dataset to train and evaluate our proposed architecture.
In~\cite{seo2018gids}, the authors present a generative adversarial network (GAN)-based IDS trained using CAN IDs and achieve an average accuracy of 97.5\% across the DoS, Fuzzy, and spoofing attacks. 
Our previous works have demonstrated that lightweight CNN-based IDS architectures, which uses received CAN IDs as its input feature, can achieve high classification accuracy across multiple attack vectors~\cite{khandelwal2022deep,khandelwal2022light}.
Recently, more complex ML architectures such as temporal convolution with global attention~\cite{cheng2022tcan}, unsupervised learning through a combination of CNN and long short-term memory (LSTM) cells~\cite{agrawal2022novelads}, gated recurrent units (GRU) networks~\cite{ma2022gru} and CNN-based temporal dependency approaches have been shown to improve detection accuracy.
These networks use CAN ID, ID $+$ Data Field, or the entire frame as the input feature as shown in table~\ref{table:inpfeatures}.
In~\cite{de2021efficient}, the authors use an iForest anomaly detection algorithm as an intrusion prevention system (IPS) to detect fuzzing and spoofing (RPM \& Gear) attacks and mark the message as an error, preventing its propagation to other ECUs; however this can cause multiple messages to be dropped from the bus in case of false positives or DoS attacks.  
In most cases, the inference models are deployed through powerful GPUs to achieve line rate detection on actual CAN bus (using CAN datasets), quantifying detection accuracy and processing latency (in batch-style operation).
Our approach aims at integrating the ML accelerators as a standard peripheral within a traditional ECU, with optimisations to achieve high accuracy and throughput, at much lower power consumption compared to traditional GPU implementations.
Hence, we quantify the detection accuracy and processing latency (in batch and per-message cases), as well as an analysis on the power consumption of the integrated approach. 

\begin{table}[ht]
\centering
\caption{Input features used by ML-based IDSs \& IPSs proposed in the research literature.}
\begin{tabular}{ll}
\toprule
\textbf{Models} & \textbf{Input Features Used}                 \\
\midrule
GIDS~\cite{seo2018gids}            & CAN ID                                       \\
DCNN~\cite{song2020vehicle}            & CAN ID                                       \\
Rec-CNN~\cite{desta2022rec}         & CAN ID                   \\
iForest~\cite{de2021efficient}         & Data Field                                   \\
MLIDS~\cite{desta2020mlids}           & CAN ID + Data Field                          \\
NovelADS~\cite{agrawal2022novelads}        & CAN ID + Data Field                          \\
TCAN-IDS~\cite{cheng2022tcan}        & CAN ID + Data Field                          \\
MTH-IDS~\cite{yang2021mth}         & CAN ID + Data Field \\
GRU~\cite{ma2022gru}               &  CAN ID + Data Field + DLC \\ 
\textbf{MLP-IDS (proposed)}         & CAN ID + Data Field \\
\bottomrule
\end{tabular}
\label{table:inpfeatures}
\end{table}

\subsection{Machine Learning on FPGAs}
While GPUs have been the target device of choice for most machine learning inference applications, FPGAs have shown great promise in recent years. 
Early efforts of deep learning accelerators on FPGAs suffered from limited on-chip memory resources, requiring most weights and biases to be stored off-chip, resulting in performance penalties compared to GPU implementations~\cite{xiao2017exploring}.
Furthermore, such accelerators had to be custom-built and optimised while also managing efficient data movement between off-chip memory and interfaces, requiring significant hardware expertise. 
Recent approaches for deploying deep learning inference on FPGAs have reduced this barrier, while still providing significant performance and energy benefits over GPUs~\cite{BLAIECH2019331, wang2016dlau}.
Tool-chains like DNN-Weaver~\cite{sharma2016dnnweaver}, LUTNET~\cite{wang2019lutnet} and FINN~\cite{umuroglu2017finn}  can efficiently map high-level representations of ML models for executing on pre-composed hardware accelerators or into fully custom accelerators. 
These tools also optimise native data representation formats (float, int)  through pruning and quantisation to reduce computational complexity and memory footprint required by the implementations.
The loss of accuracy in such optimisations is addressed by augmenting training steps and/or post-quantisation fine-tuning. 
With some hardware expertise, these frameworks will allow every layer of the network to be optimised (including weights, biases, and activation functions) and right-sized  (down to 1-bit in the case of FINN) with minimal impact on accuracy, generating energy-efficient high-performance implementations, which can then be connected as slave accelerators in an embedded design or used as standalone accelerators. 
Alternatively, vendor flows like Vitis-AI from Xilinx~\cite{xilinxvitis} map a pre-trained network onto a selection of pre-composed configurable deep-learning processing units (DPUs)~\cite{xilinxdpu} on the hardware, lowering the barrier for using FPGA-based ML accelerators.
The Vitis-AI flow quantizes and (optionally) fine-tunes the pre-trained model to generate the hardware accelerator model along with its drivers and APIs for offloading computations to it from a host code. 
While not fully optimised as a custom implementation generated through FINN, the Vitis-AI flow offers a compelling alternative to the GPU flow without a significant drop in energy efficiency compared to fully custom networks at this precision (int8 precision), and hence, in this paper, we utilise the Vitis-AI flow to design and deploy our IDS accelerator on the hybrid Zynq Ultrascale+ device. 

FPGA-based deployments are gaining momentum in the automotive domain, with prior research having explored acceleration of real-time complex algorithms for vision-based advanced driver assistance systems (ADAS) and using network interface extensions to augment ECU functionality and security~\cite{shreejith2018smart}.
FPGA-based ECUs architectures have also been explored to enable custom capabilities while maintaining complete AUTOSAR compliance~\cite{fons2012fpga}.
Automotive suppliers and vendors have also explored hybrid FPGA-based ECUs for improved functional consolidation and reliability~\cite{BoschMS6}. 
Specifically for network security, ML-based anomaly detection (bit-timing~\cite{zhou2019btmonitor}, transmission timing~\cite{yang2020identify}) models, probabilistic attack models~\cite{8987605} and generalised multi-attack detection models~\cite{khandelwal2022deep,khandelwal2022light} have leveraged FPGA platforms for improved latency and detection performance.  
With the rising computational complexity in vehicles, it is expected that hybrid FPGA based ECUs will gain more traction for complex vehicular functions. 
Our approach aims at a integrated IDS-ECU architecture which consolidates the intrusion detection function through dedicated slave accelerator(s) with the primary ECU function on the same chip/package, mimicking an off-the-shelf ECU architecture.
The accelerators will execute our QMLP models directly off received CAN messages from the network in full isolation and we show that such integration offers unique benefits in terms of latency, energy efficiency and detection accuracy across multiple attack vectors. 
\section{System Architecture}\label{sec:proposedmodel}
\subsection{IDS-ECU Architecture}
\begin{figure}[t!]
    \centering
    \includegraphics[scale = 0.72]{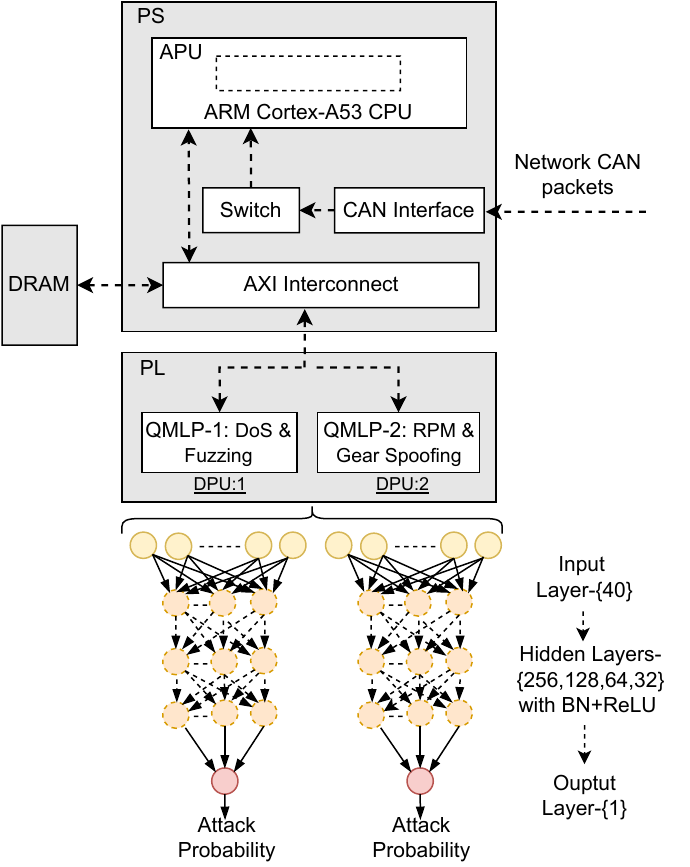}
    \caption{Proposed system architecture of the IDS-ECU with the quantised MLP accelerators on the PL part of the FPGA.}
    \label{fig:datapath}
\end{figure}

Figure~\ref{fig:datapath} shows the proposed ECU architecture of the IDS-enabled ECU on a Xilinx Zynq Ultrascale+ device. 
The Zynq device integrate capable ARM processors connected to a host of hardened memory mapped peripheral logic and interface protocols within the processing system (PS) section of the device. 
Any custom peripheral can be integrated into the programmable logic (PL) region, and wired with the PS using high or low performance Advaned eXtensible Interface (AXI) ports and can be accessed as memory mapped devices from the software application on the processor.
In our proposed ECU architecture, standard ECU function(s) are mapped as software tasks onto one or more of the ARM cores on the PS, on top of a standard operating system like Linux or a real-time operating system. 
The operating system provides relevant drivers and APIs for accessing the PS peripherals and the PL accelerators, abstracting away low-level details of these blocks to create an AUTOSAR compliant architecture~\cite{fons2012fpga}.
We propose to use the integrated CAN interface on the PS to handle the interfacing of ECU to the CAN bus, as shown in Figure~\ref{fig:datapath}.

The PL instantiates the Xilinx DPU~\cite{xilinxdpu} block as our dedicated ML accelerator which can be configured and controlled using the Vitis AI Runtime (VART) APIs from the PS. 
The standard Vitis-AI flow is used to automatically generate the DPU IP along with wrappers, interconnects and the runtime elements from our high-level ML model described in TensorFlow (see Sec.~\ref{subsec:MLmodel}).
When the CAN interface receives a valid CAN packet, the software task on the PS reads the message into its buffer (RAM) for the ECU application to process and take appropriate actions according to the tasks' specification.
The isolated IDS task adds the relevant information from the CAN packet to the input feature buffer and uses the integrated Vitis-AI Runtime (VART) APIs to activate the IDS.
The input feature buffer is organised as a FIFO (depth of 4 messages) to capture the temporal features of adjacent messages on the CAN bus. 
We integrate two DPU cores which execute in concurrent fashion (non-blocking execution mode), allowing the input features to be evaluated for multiple threat vectors simultaneously.  
The DPU cores then run the model on the packet information and interrupt the PS with a completion status.  
This scheme allows for a seamless integration with the normal event/time-triggered tasks to be executed by the ECU functions.

\subsection{MLP-based IDS}
\label{subsec:MLmodel}
To determine the best ML model, we profiled different network architectures with varying complexity to find a baseline model that offers high inference accuracy at minimal complexity. 
We observed that while CNNs were effective for classification (with CAN IDs as input features), the MLP design provided more accurate detection (accuracy, false positives, and false negatives) when using the entire message contents at much lower computational complexity, and was hence chosen as our choice of architecture.  
A concatenation of \textit{n = 4} successive messages (CAN IDs + payload) is also chosen as the input feature based on this testing. 
Our model exploration was also guided by the network layers supported by the Vitis-AI framework. 
We used the open Car Hacking dataset for Intrusion Detection for our profiling, training, and validation~\cite{song2020vehicle}.

The model consists of 5 \emph{Dense} layers implemented with the 256, 128, 64, 32 \& 1 units at each layer.
The time-series data from the input feature buffer is fed as input to the input layer with 40 units. 
Subsequent layer(s) operate on the output of the previous layer, with increasing complexity to perform classification.
Batch Normalisation and Dropout layers were used between the dense layers to prevent over-fitting and to improve the learning efficiency during the training phase. 
The output layer uses a \emph{sigmoid} function at the output of the final dense layer to estimate the probability of an infected message. 
The model is defined in TensorFlow (TF) using standard TF functions and nodes.

\subsection{Dataset and Training}
\label{subsec:dataset}
As mentioned above, we use the open Car Hacking dataset for training our model and to test its performance~\cite{canlink}. 
The dataset provides a labelled set of normal and attack messages which were captured via the Onboard Diagnostic (OBD) port in an actual vehicle, with attack messages injected in real-time. 
The dataset includes DoS, Fuzzing and Spoofing message injections allowing us to validate the detection accuracy across these different attacks. 
An extract from the dataset is shown in table~\ref{table:dataset}, showing the CAN ID (ID field), control field [Data Length Code (DLC) field] and the actual data segment (Data field).
We split the dataset as 80:15:5 for training, validation, and testing respectively, allocating the large section to training and optimisation of the quantised network. 
The performance of the model on the validation set during training ensures that it is not over-fitting on any of the attacks. 
We pre-process the dataset prior to training to mimic the dataflow the model will obtain as its input when integrated into the ECU. 
The message fields (ID and payload) of each message are packed into INT8 type vectors, and a sequence of \textit{n = 4} of these messages form the input feature buffer content for the IDS (block shape of \textit{\{1,10*n\}} INT8 data).
Each layer of this stack forms the combined input shape for the first layer, which is reshaped to feed into the exact channels for training and testing.

\begin{table}[t!]
\centering
\caption{An extract from the open Car hacking dataset which is used for our testing and evaluation.}
\scalebox{1}{
\begin{tabular}{@{}lrrr@{}}
\toprule
\textbf{Time}     & \textbf{ID} & \textbf{DLC} & \textbf{Data}           \\ \midrule
\textbf{$\ldots$} \\
1478198376.389427 & 0316            & 8            & 05,21,68,09,21,21,00,6f \\
1478198376.389636 & 018f            & 8            & fe,5b,00,00,00,3c,00,00 \\
1478198376.389864 & 0260            & 8            & 19,21,22,30,08,8e,6d,3a \\ 
\textbf{$\ldots$} \\
\bottomrule
\end{tabular}}
\label{table:dataset}\vspace{-5mm}
\end{table}

To train the model, we used adam optimizer with a binary cross-entropy loss function. 
The learning rate was set to 0.0001 to allow for slower learning which aids in reducing loss of accuracy when quantising the pre-trained model~\cite{wu2018training}.
For the first MLP, we first train the model on DoS attack and ensure optimal performance; subsequently, this trained model is trained on the fuzzing dataset to improve generalisation across the two attack modes through inductive transfer. 
This model file is then tested in both the DoS \& Fuzzing attacks to ensure there was no performance degradation.
The model was trained for 50 epochs with a batch size of 64.
The model saves intermediate results at each epoch allowing us to progressively track and integrate early stopping in case of a significant drop in the accuracy. 
The same training \& validation flow is followed for the second MLP that is trained to classify RPM \& Gear spoofing attacks.
The training was performed on a workstation class machine with an Intel i9-9820X and an Nvidia A6000 GPU.
The trained model files were exported as a `.h5' checkpoint for optimisation by the Vitis-AI flow.

\subsection{Mapping to DPU and Integration}
The trained checkpoint files in full floating-point representation are fed as the input to the Vitis-AI framework along with the training and validation sets from the dataset. 
The framework executes multiple optimisation passes, quantising the weights, activation, and biases at each pass and then evaluating the resulting accuracy of the resultant model to determine the steps for the next pass.
This flow converts all parameters and operations to 8-bit (INT8) format, with minimal loss in inference accuracy.
We further fine-tune the quantised model through a quantisation-aware optimisation step which improved the inference accuracy of the quantised model, particularly for the first MLP detecting DoS \& Fuzzing attacks.
Once optimized, the Vitis flow compiles the model to a DPU-executable `xmodel' file, which incorporates the (quantised) weights, biases, and instruction sequences for executing the model on the DPU. 
Simultaneously, it also generates a hardware project with the DPU integrated as a slave peripheral of the Zynq PS along with the run-time libraries and API wrappers for the Linux operating system.
Each DPU uses three master and one slave AXI interface to move instructions and data (inputs/weights/biases) between the PS and its internal memory to help maximise the peak performance. 
For this implementation, two instances of the B512 DPU configuration is deployed to minimise the resource overhead and power consumption. 
The Vivado flow generates the boot image for the Zynq device for booting from an external SD card with a standard Linux kernel.

\section{Deployment and Experimental Results}\label{sec:experiments}

For training, we use the floating point variants of both models and use an Nvidia A6000 GPU for accelerating the training and validation, prior to quantisation through Vitis-AI.
Post quantisation, the lightweight models have 53,300 parameters each and are packed into the `xmodel' executable format. 
For our test, we use a Zynq Ultrascale+ ZCU104 development board which features a XCZU7EV Ultrascale+ device with quad-core ARM A53 cores and dual-core ARM R5 cores on the PS.
A standard Linux kernel with petalinux tools and VART interfaces enabled is used as the boot configuration for the ARM cores.
%
The VART APIs are used to deploy the models to the DPUs (in PL) at startup and to invoke test cases at run time.
The A53 cores on the PS are configured to run at 1.2\,GHz peak. 
The DPUs use a 600\,MHz DSP core clock for its execution and are configured to run concurrently for all our tests at startup. 
The Nvidia A6000 GPU has an base operating frequency of 1350\,MHz.

We quantify the accuracy of inference by evaluating precision, recall, F1 rates as well as the false positive and false negative rates (FPR and FNR respectively) for our pre-quantised and quantised variants (on GPU and FPGA respectively).
The processing latency and power consumption for performing IDS on each incoming CAN message on the ECU (using the DPU accelerators) is measured and compared against the inference on the A6000 GPU (mimicking a loosely-coupled accelerator) to compare the inference time/power required for each new CAN message on the two platforms.
The inference accuracy metrics and message processing latency are compared against state-of-the-art IDSs/IPSs proposed in the literature.
In case of schemes where inference is performed on a block of CAN messages, we use these metrics along with the block size for the comparison.
We also compare our active power consumption against ML-IDS approaches in literature where power consumption has been reported. 

\subsection{Accuracy}
To test the functional correctness of the models, we first compare the pre- \& post-quantisation inference performance of both models. 
We see a drop in the performance of QMLP-1 on the fuzzing attack which is corrected by the Vitis-AI post quantisation training flow as shown in table~\ref{table:mlp1_perf}.
Table~\ref{table:mlp2_perf} shows the performance of QMLP-2 on the spoofing attacks for which we see no such degradation.  

\begin{table*}[!t]
\centering
\caption{Inference accuracy metrics of QMLP-1, pre and post quantisation \{Before and After Fine-tuning (BF \& AF respectively)\} on the DoS and Fuzzing attacks.}
\begin{tabular}{c ccc ccc ccr ccr ccr}
\toprule
& \multicolumn{3}{c}{\textbf{Precision}} & \multicolumn{3}{c}{\textbf{Recall}}   & \multicolumn{3}{c}{\textbf{F1-Score}} & \multicolumn{3}{c}{\textbf{FPR}} & \multicolumn{3}{c}{\textbf{FNR}} \\ \midrule
                 & \multicolumn{1}{c}{} & \multicolumn{2}{c}{Post-Q}  & \multicolumn{1}{c}{}     & \multicolumn{2}{c}{Post-Q}     & \multicolumn{1}{c}{}  & \multicolumn{2}{c}{Post-Q}  & \multicolumn{1}{c}{}    & \multicolumn{2}{c}{Post-Q}  & \multicolumn{1}{c}{}                        & \multicolumn{2}{c}{Post-Q}                         \\  \cmidrule{3-4} \cmidrule{6-7} \cmidrule{9-10} \cmidrule{12-13} \cmidrule{15-16} 
                 & \multicolumn{1}{c}{\multirow{-2}{*}{Pre-Q}}                              & \multicolumn{1}{c}{BF}                                                   & AF                                                            & \multicolumn{1}{c}{\multirow{-2}{*}{Pre-Q}}                              & \multicolumn{1}{c}{BF}                                                   & AF                                                            & \multicolumn{1}{c}{\multirow{-2}{*}{Pre-Q}}                              & \multicolumn{1}{c}{BF}    & \multicolumn{1}{c}{AF} & \multicolumn{1}{c}{\multirow{-2}{*}{Pre-Q}} & \multicolumn{1}{c}{BF}   & \multicolumn{1}{c}{AF} & \multicolumn{1}{c}{\multirow{-2}{*}{Pre-Q}} & \multicolumn{1}{c}{BF}   & \multicolumn{1}{c}{AF} \\ \midrule
DoS     & \multicolumn{1}{c}{ 99.81} & \multicolumn{1}{c}{ 99.58} &  \textbf{99.92} & \multicolumn{1}{c}{ 100}   & \multicolumn{1}{c}{ 100} &  \textbf{100} & \multicolumn{1}{c}{ 99.91} & \multicolumn{1}{r}{99.79} & \textbf{99.96}          & \multicolumn{1}{r}{0.09}                    & \multicolumn{1}{r}{0.21} & \textbf{0.04}           & \multicolumn{1}{r}{0}                       & \multicolumn{1}{r}{0} & \textbf{0}           \\ 
Fuzzing & \multicolumn{1}{c}{ 99.85} & \multicolumn{1}{c}{91.08} & \textbf{99.86} & \multicolumn{1}{c}{98.15} & \multicolumn{1}{c}{ 99.25} &  \textbf{99.67} & \multicolumn{1}{c}{ 98.99} & \multicolumn{1}{r}{94.99} & \textbf{99.76}          & \multicolumn{1}{r}{0.04}                    & \multicolumn{1}{r}{2.8}  & \textbf{0.04}           & \multicolumn{1}{r}{1.85}                    & \multicolumn{1}{r}{0.75} & \textbf{0.33}           \\ \bottomrule
\end{tabular}
\label{table:mlp1_perf}
\end{table*}

\begin{table}[t!]
\centering
\caption{Inference accuracy metrics of QMLP-2 pre and post quantisation on the RPM \& Gear attacks.} 
\scalebox{1}{
\begin{tabular}{@{}lllllll@{}}
\toprule
\textbf{Attack} & \textbf{Model}  & \textbf{Precision} & \textbf{Recall} & \textbf{F1} & \textbf{FPR} & \textbf{FNR} \\
\midrule
\multirow{2}{*}{RPM} & Pre-Q          & 100             & 100               & 100    &0  & 0  \\
& QMLP-2         & 100             & 100               & 100    & 0 &  0 \\
\multirow{2}{*}{Gear} & Pre-Q          & 100             & 100          & 100    &  0& 0  \\
& QMLP-2         & 99.90             & 100          & 99.95   &0.02  &   0 \\ 
\bottomrule
\end{tabular}}
\label{table:mlp2_perf}
\end{table}

We compare the inference performance of our QMLPs integrated within the ECU against the state-of-the-art IDSs and IPSs proposed in the literature: GIDS~\cite{seo2018gids}, DCNN~\cite{song2020vehicle}, MLIDS~\cite{desta2020mlids}, NovelADS~\cite{agrawal2022novelads}, TCAN-IDS~\cite{cheng2022tcan}, IForest~\cite{de2021efficient}, MTH-IDS~\cite{yang2021mth}, GRU~\cite{ma2022gru} and Rec-CNN~\cite{desta2022rec}.
F1 score, precision, and recall metrics are used for comparison since most models report the inference performance using these metrics.
The detailed results are tabulated in tables~\ref{table:comp1} and ~\ref{table:comp2} for DoS-Fuzzy attacks and Spoofing attacks respectively. 
Both MTH-IDS~\cite{yang2021mth} and Rec-CNN~\cite{desta2022rec} report an average accuracy, precision \& recall of 99.9\% across all attacks which is identical to our model on each of the three metrics. 
In GRU~\cite{ma2022gru}, authors report average precision, recall and F1 score values of 99.95\%, 99.31\% \& 99.63\% respectively for the spoofing attacks. 
Our model achieves the same (MTH-IDS \& Rec-CNN) or better (GRU) results on all three metrics for these attacks.
 
From table~\ref{table:comp1}, it can be observed that in case of the \textit{DoS} attack, our approach offers identical performance to DCNN, MLIDS, NovelADS, TCAN-IDS and GRU models (F1 score).
For \textit{fuzzing attack}, our approach offers identical inference accuracy compared to DCNN, MLIDS, NovelADS, TCAN-IDS and performs better than iForest and GRU models by 2.3\% and 0.5\%, respectively (F1 score).
Similarly, from table~\ref{table:comp2}, it can be observed that our approach achieves $\ge$ 99.95\% performance across both spoofing attacks across all parameters, and also perform identical to or better than others reported in the literature. 
It should also be noted that many reported results in the literature rely on retraining the model for each attack and measuring the inference accuracy independently and hence require multiple models to be deployed concurrently to detect all the attack vectors~\cite{song2020vehicle,desta2022rec,agrawal2022novelads,cheng2022tcan,ma2022gru}. 
In contrast, our approach uses a single integrated IDS-ECU and transfer learning during the training phase to allow every message to be simultaneously evaluated for these different attack signatures.
%

\begin{figure}[t]
\centering
\begin{tabular}{ll}
\hspace*{-0.5cm} 
\includegraphics[scale=0.3,trim={0 0.5cm 0 2cm},clip]{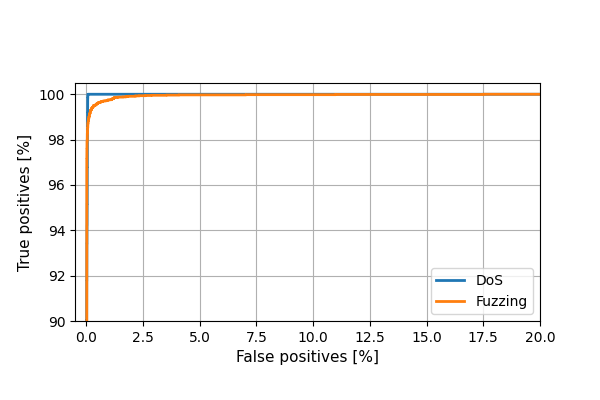} &
\hspace{-0.5cm} 
\includegraphics[scale=0.3,trim={0 0.5cm 0 2cm},clip]{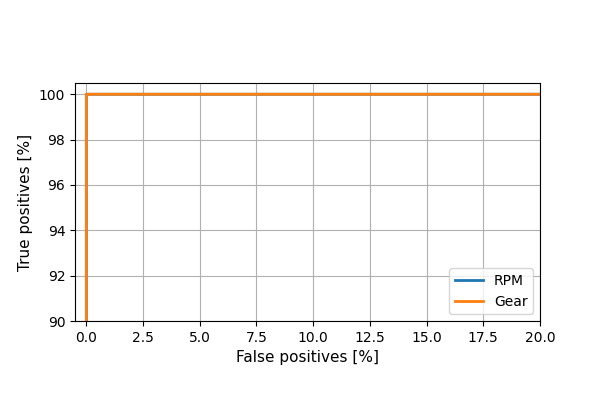}
\end{tabular}
\caption{ROC curve of the MA-QMLPs on the four attacks.}
\label{fig:roc}\vspace{-5mm}
\end{figure}

Tables~\ref{table:confmatrix_mlp1} \&~\ref{table:confmatrix_mlp2} captures the confusion matrix of our QMLP models on the four attacks across the test vectors, showing the raw classification numbers achieved in each case.
The misclassifications in our test set can be attributed to scenarios with nearly identical attack and normal message patterns among a stream of messages and/or within the observed block of messages at a given time.
Figure~\ref{fig:roc} shows the receiver operating characteristic (ROC) curves of the two QMLP models across the test vectors indicating that our classifier performs well for all discrimination thresholds. 
From the ROC curve, we can also observe that the area under curve (AUC) values achieved is $\ge$ 0.99 for all four attacks, showing that the model has a high probability of correctly distinguishing between normal and attack messages.

\begin{table}[t!]
\centering
\caption{Accuracy metric comparison (\%) of our quantised FPGA accelerator (QMLP-1) against the reported literature on the DoS and Fuzzing attacks.}
\scalebox{1}{
\begin{tabular}{@{}llllll@{}}
\toprule
\textbf{Attack}  & \textbf{Model} & \textbf{Precision} & \textbf{Recall} & \textbf{F1}  & \textbf{FNR} \\
\midrule
\multirow{5}{*}{DoS} & GIDS~\cite{seo2018gids}                  & -                & 99.9          & -  &   -   \\ 
& DCNN~\cite{song2020vehicle}                  & 100                & 99.89          & 99.95  & 0.13     \\
& MLIDS~\cite{desta2020mlids}                  & 99.9                & 100          & 99.9  & -     \\
& NovelADS~\cite{agrawal2022novelads}                  & 99.97               & 99.91          & 99.94  & -     \\
& TCAN-IDS~\cite{cheng2022tcan}                  & 100             & 99.97          & 99.98  & -     \\
& iForest~\cite{de2021efficient}                  & -                &   -       &  - &  -    \\ 
& GRU~\cite{ma2022gru}                  & 99.93             & 99.91               & 99.92  & -     \\
& \textbf{QMLP-1 (DPU)}                  & 99.92             & 100               & 99.96  & 0     \\
\midrule
\multirow{5}{*}{Fuzzy} & GIDS~\cite{seo2018gids}                  & -                & 98.7          & -   & -    \\ 
& DCNN~\cite{song2020vehicle}                 & 99.95             & 99.65          & 99.80  & 0.5     \\
& MLIDS~\cite{desta2020mlids}                  & 99.9             & 99.9          & 99.9  & -     \\
& NovelADS~\cite{agrawal2022novelads}                  & 99.99               & 100         & 100  & -     \\
& TCAN-IDS~\cite{cheng2022tcan}                  & 99.96             & 99.89          & 99.22  & -     \\
& iForest~\cite{de2021efficient}                  & 95.07                & 99.93          & 97.44  &    -  \\
& GRU~\cite{ma2022gru}                  & 99.32             & 99.13               & 99.22  & -     \\
& \textbf{QMLP-1 (DPU)}            & 99.86             & 99.67          & 99.76   &  0.33    \\
\bottomrule
\end{tabular}}
\label{table:comp1}
\end{table}

\begin{table}[t]
\centering
\caption{Accuracy metric comparison (\%) of our quantised FPGA accelerator (QMLP-2) against the reported literature on the RPM and Gear attacks.}
\scalebox{1}{
\begin{tabular}{@{}llllll@{}}
\toprule
\textbf{Attack}  & \textbf{Model} & \textbf{Precision} & \textbf{Recall} & \textbf{F1}  & \textbf{FNR} \\
\midrule
\multirow{5}{*}{RPM} & GIDS~\cite{seo2018gids}                  &     -           & 99.6          & -  &   -   \\ 
& DCNN~\cite{song2020vehicle}                  &  99.9               &     99.9      & 99.9   & 0.05     \\
& MLIDS~\cite{desta2020mlids}                  &    100             &   100        & 100  & -     \\
& NovelADS~\cite{agrawal2022novelads}                  &  99.9             &   99.9        & 99.9 & -     \\
& TCAN-IDS~\cite{cheng2022tcan}                  &     99.9         &     99.9      &  99.9 & -     \\
& iForest~\cite{de2021efficient}                  & 98.9               &   100       &  99.4 &  -    \\ 
& \textbf{QMLP-2 (DPU)}                  &   100           & 100               & 100  & 0     \\
\midrule
\multirow{5}{*}{Gear} & GIDS~\cite{seo2018gids}                  &   -              &   99.8        &  -  & -    \\ 
& DCNN~\cite{song2020vehicle}                 &     99.9         &   99.8        &  99.9 &   0.11   \\
& MLIDS~\cite{desta2020mlids}                  &     100         &    100       &  100 & -     \\
& NovelADS~\cite{agrawal2022novelads}                  &    99.8            &   99.9       & 99.9  & -     \\
& TCAN-IDS~\cite{cheng2022tcan}                  &  99.9           &     99.9      & 99.9  & -     \\
& iForest~\cite{de2021efficient}                  &       94.7          &   100        &  97.3 &    -  \\
& \textbf{QMLP-2 (DPU)}            & 99.9             & 100          & 99.95   & 0     \\
\bottomrule
\end{tabular}}
\label{table:comp2}
\end{table}

\begin{table}[t!]
\centering
\caption{Confusion matrix capturing the classification results of our QMLP-1 on the DoS and Fuzzing attacks.}
    \scalebox{1}{
        \begin{tabular}{@{}llrr@{}}
        \toprule
            \textbf{Attack} & \textbf{Message Type}  & Predicted Normal  & Predicted Attack  \\ 
            \midrule
          \multirow{2}{*}{DoS} & True Normal    &  33282                                & 13                                   \\ 
            & True Attack    & 0                                 & 16705                                \\ 
            \multirow{2}{*}{Fuzzy} & True Normal    &    38806                            &   16                                 \\
           &  True Attack    &    37                              &   11141                              \\ 
            \bottomrule
        \end{tabular}}
\label{table:confmatrix_mlp1}
\end{table}

\begin{table}[t!]
\centering
\caption{Confusion matrix capturing the classification results of our QMLP-2 on the RPM and Gear spoofing attacks.}
    \scalebox{1}{
        \begin{tabular}{@{}llrr@{}}
        \toprule
            \textbf{Attack} & \textbf{Message Type}  & Predicted Normal  & Predicted Attack  \\ 
            \midrule
          \multirow{2}{*}{RPM} & True Normal    &  40221                                & 0                                   \\ 
            & True Attack    & 0                                 & 9779                                \\ 
            \multirow{2}{*}{Gear} & True Normal    &    41352                            &   9                                 \\
           &  True Attack    &    0                              &   8639                              \\ 
            \bottomrule
        \end{tabular}}
\label{table:confmatrix_mlp2}
\end{table}

\subsection{Inference latency}
We quantify the per-message processing latency to show the potential of a tightly coupled IDS with ECU functionality to perform IDS on the message as soon as they arrive. 
The latency measure includes the time for move data from the message buffer on the PS to the PL, execution time and to read back the results to the software task on the PS. 
Note that the PS task that initiates the IDS operation is non-blocking, allowing the processor to do other tasks during the IDS execution. 
The latency measurement is compared against others reported in the literature in table~\ref{table:latcomp}, providing a comparison against approaches which run different models per message or per block of messages and on a variety of platforms as isolated IDS and loosely coupled accelerators. 
The tightly coupled IDS in our ECU model achieve an end-to-end message processing latency of 0.24\,ms for both the models since they are executed concurrently. 
This is a 2.3$\times$ speed up compared to MTH-IDS implemented on a dedicated IDS-ECU on a Raspberry-Pi 3 device.
We also observe a significant speed-up compared to a loosely coupled GPU architecture performing inference (on the Nvidia A6000) as shown in table~\ref{table:lat_power}; a portion of the higher latency can be attributed to the PCIe overheads in small data size transactions in a per-message execution mode. 
In terms of raw throughput, the tightly coupled IDS-ECU can process 4166 messages per second scanning for all four attacks.
This translates to a line-rate detection at 462\,Kbps on a high-speed CAN network (at max data size) using the lightweight DPU. 
While GPU-based methods have reported better average throughput by processing a block of CAN messages at once, the detection delay incurred by the time taken to accumulate the block size of messages is ignored in such calculation. 
For instance, accumulating 64 CAN messages at peak data rate (1\,Mbps) takes 8.3\,ms, which is not factored into the reported detection latency of 5.89\,ms in the case of GIDS in table~\ref{table:latcomp}.

We also quantify the cold-start setup time of IDS to determine the delay from power on to the model being ready to perform IDS.  
This is an important consideration since many attacks could be initiated during the startup phase of the network and ECUs.
In the case of our test platform, we observe this cold-start latency as 8.9\,seconds when averaged across multiple runs. 
It should be noted that our test platform uses a slower boot device (SD card interface) and full Linux OS, whereas typical ECUs rely on non-volatile flash memory and real-time operating system, which could reduce the cold-start latency by 8$\times$ (with Linux and flash boot) or more. 

\begin{table}[t]
\centering
\caption{Per-message latency comparison against other state-of-the-art IDSs reported in literature.}
\scalebox{1}{
\begin{tabular}{@{}lrll@{}}
\toprule
\textbf{Models}     & \textbf{Latency} & \textbf{Frames} & \textbf{Platform} \\ \cmidrule{1-4}
GRU~\cite{ma2022gru} & 890\,ms & 5000 CAN frames & Jetson Xavier NX  \\
MLIDS~\cite{desta2020mlids} & 275\,ms & per CAN frame & GTX Titan X \\
Rec-CNN~\cite{desta2022rec} & 117\,ms & 128 CAN frames & Jetson TX2 \\
NovelADS~\cite{agrawal2022novelads} & 128.7\,ms & 100 CAN frames & Jetson Nano \\
GIDS~\cite{seo2018gids} & 5.89\,ms & 64 CAN frames &  GTX 1080  \\
DCNN~\cite{song2020vehicle} &  5\,ms & 29 CAN frames & Tesla K80 \\
TCAN-IDS~\cite{cheng2022tcan} & 3.4\,ms & 64 CAN frames & Jetson AGX \\
MTH-IDS~\cite{yang2021mth} &  0.574\,ms & per CAN frame & Raspberry Pi 3     \\
\textbf{QMLP-IDS} (ours) & 0.24 ms & per CAN frame & Zynq Ultrascale+    \\ 
\bottomrule
\end{tabular}}
\label{table:latcomp}
\end{table}

\subsection{Power consumption, PS/PL utilisation}
We quantify the power consumption of our IDS-ECU using the PYNQ-PMBus package to monitor the power rails directly and average them over 100 runs.
At idle, with the ECU booted up with the Linux image, we observe an average consumption of 3.43\,W from the power rails, which rises to 3.76\,W when CAN messages are passed to the IDS for execution.  
Considering the total power consumed by the device, this leads to the average power consumption of 0.9mJ per inference. 
To model a loosely coupled GPU accelerator, we measure the power consumption of the quantised model using an A6000 GPU and the Nvidia 'nvtop' plug-in (on a workstation machine) and observe an active power consumption of 56W (GPU only) incurring 1.96\,J per inference. 
The IDS-ECU approach hence achieves a 15$\times$ reduction in power consumption as shown in table~\ref{table:lat_power} and a 2172$\times$ better energy efficiency per inference making our approach a more efficient architecture for integrating IDS capabilities across ECUs. 
Among other reported results in the literature, our approach incurs a 2.6$\times$ reduction in power consumption when compared to the GRU~\cite{ma2022gru} model on an Nvidia Jetson Xavier as a dedicated IDS node.

\begin{table}[t!]
\centering
\caption{Per-message latency and measured power on the different platforms.}
\begin{tabular}{@{}lrrr@{}}
\toprule
\multirow{2}{*}{\textbf{Platform}} & \textbf{Latency} & \multicolumn{2}{c}{\textbf{Measured Power}} \\ \cmidrule{3-4}
    & (ms) & Idle (W) & Active (W) \\ \midrule
QMLP on RTX A6000  &  35 & 22  & 56  \\
\textbf{QMLP on IDS-ECU} & 0.24  & 3.43  & 3.76  \\
\bottomrule
\end{tabular}
\label{table:lat_power}
\end{table}

Finally, we quantify the overall hardware resource consumption of the IDS in table~\ref{table:resourceutilization} and observe a peak consumption of 27.8\% of resources for BRAMs and 24.6\% for LUTs. 
This leaves enough resources ($>$ 75\% LUTs \& $>$ 85\% DSP blocks) on the PL to allow other ECU functionality to be offloaded for hardware acceleration.
We also quantify the utilisation of the ARM cores for managing the IDS accelerator to estimate the overhead in consolidating the IDS capability on the ECU. 
It was observed that a single core utilisation peaked at 40\% when processing the completion interrupt from the IDS core (for each message at near-line-rate), while other cores remained at IDLE ($\le 2\% $ utilisation). 
We believe that this could be further refined by processing the IDS outputs in hardware to flag the threats to the ECU, while also improving the detection latency.  

\begin{table}[t!]
\centering
\caption{Resource utilisation on the PL for the proposed CAN IDS-ECU (XCZU7EV).}
\scalebox{1}{
\begin{tabular}{lrrrr}
\toprule
\textbf{Node} & \textbf{LUT} & \textbf{FF} & \textbf{BRAM/URAM} & \textbf{DSP} \\
\midrule
DPU:Core-1           & 27020        &  33889           & 41.5/12            & 110  \\
DPU:Core-2           & 27018        &  33953           & 41.5/12            & 110  \\
\midrule
Overall & 56733  & 72147  & 87/24  & 220 \\
(\% usage) & (24.6\%) & (15.6\%) & (27.8\%/25\%) & (12.73\%) \\
\bottomrule
\end{tabular}}
\label{table:resourceutilization}
\end{table}

\section{Conclusion}\label{sec:conclusion}
In this paper, we present an ECU architecture that consolidates IDS capabilities for CAN interface through a dedicated multi-core accelerator. 
The tightly coupled IDS-ECU architecture is capable of detecting DoS, Fuzzing \& spoofing attacks using two lightweight quantised MLPs and deployed using two resource-efficient variants of Xilinx's DPU accelerator.
Our lightweight MLP model matches or improves upon the detection accuracy metrics across multiple attack vectors using a single deployment, compared to the state-of-the-art models which require retraining and tuning to each attack vector. 
The proposed integration achieves a 2.3$\times$ reduction in per-message intrusion detection latency and a 2.6$\times$ reduction in power consumption compared to the state-of-the-art IDSs proposed in the literature that uses standalone and/or loosely coupled IDS integration on CPUs/GPUs.
Our results show that the proposed architecture is best suited for the deployment of consolidated IDS capabilities within critical ECUs in a distributed fashion within the CAN network. 
In the future, we propose to expand our approach to integrate emerging vehicular networks like Automotive Ethernet and integration at network gateways for next-generation vehicular architectures.
\section{Acknowledgement}
This research was supported by grants from NVIDIA and utilised NVIDIA RTX A6000 GPU.
\bibliography{references}
\bibliographystyle{ieeetr}

\end{document}